\documentclass[twoside,a4paper,11pt]{sea22}
\usepackage{graphicx}
\usepackage{hyperref}
\usepackage{media9}
\begin{document}
\pagenumbering{arabic}
\pagestyle{myheadings}
\thispagestyle{empty}
{\flushleft\includegraphics[width=\textwidth,bb=58 650 590 680]{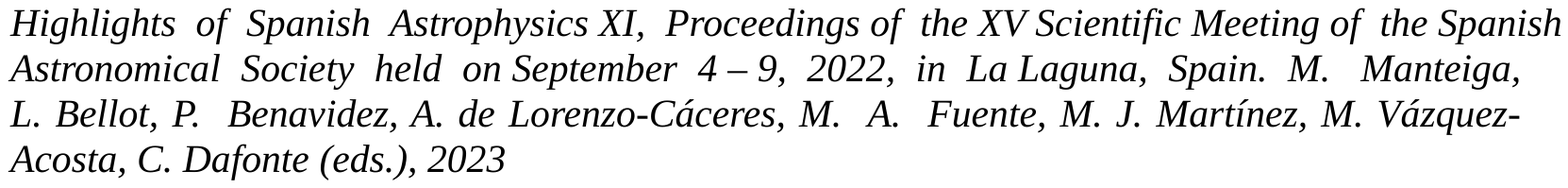}}
\vspace*{0.2cm}
\begin{flushleft}
{\bf {\LARGE
%
A tactile model of the night summer northern sky for the teaching of astronomy to the BVI
%
}\\
\vspace*{1cm}
%
P\'erez-Montero, E.$^{1}$,
Lanzara, M.$^{2}$, 
Ortiz-Gil, A.$^{2}$,
Villaverde, M.$^{1}$,
Garc\'\i a-Benito, R.$^{1}$,
Gallego-Calvente, T.$^{1}$,
and 
Garc\'\i a G\'omez-Caro, E.$^{1}$
%
}\\
\vspace*{0.5cm}
%
$^{1}$
Instituto de Astrof\'\i sica de Andaluc\'\i a - CSIC. Apdo. de correos 3004, E-18080, Granada (Spain) \\
$^{2}$
University of Valencia, Astronomical Observatory, Paterna (Valencia), Spain\\
%
\end{flushleft}
%
\markboth{
A tactile model of the summer night sky
}{ 
%
P\'erez-Montero et al. 
%
}
\thispagestyle{empty}
\vspace*{0.4cm}
\begin{minipage}[l]{0.09\textwidth}
\ 
\end{minipage}
\begin{minipage}[r]{0.9\textwidth}
\vspace{1cm}
\section*{Abstract}{\small
%
{\em Astroaccesible} is an outreach project hosted by the Instituto de Astrof\'\i sica de Andaluc\'\i a - CSIC aimed at the teaching and popularisation of the astronomy among all publics independently of their capabilities and abilities, paying special attention to the collective of blind and visually impaired (BVI).
Among the different strategies and resources using in our project, we have developed new 3D models representing in relief some of the stars, constellations and deep sky objects that can be observed during night from
the Northern hemisphere in spring and summer. These models can be used by BVI to transmit to them the spatial configuration of the sky during night, but can be also used as an additional
resource for all kind of publics to complement their sensorial experience. We also describe additional resources based on sounds that can also be employed to get deeper into this multisensorial experience. Finally, we summarize some of the activities and the context in which this new material has been used in the last 2 years. 
%
\normalsize}
\end{minipage}
%
%
%
\section{Introduction \label{intro}}
The outreach project {\em Astroaccesible}\footnote{\href{http://astroaccesible.iaa.es}{http://astroaccesible.iaa.es}} \cite{pm16} has the objective of bringing astronomy to every public, using all kind of
resources based on the senses other than that of sight in order to reach to the collective of blind and visually impaired (BVI). The objective of using this material and strategies for every public, independently of their abilities, is fundamented on that  many of these complementary resources  strength and deepen the attention and the clarity of the transmitted contents \cite{pm19}. Therefore, {\em Astroaccesible} has also the objective of convincing other scientists and teaching professionals to incorporate an inclusive aspect to their projects, as this largely benefit the collective of impaired people, helping at the same time to improve the quality of their contents for everyone.

Among the different activities carried out by {\em Astroaccesible} in the last years there are thus sessions devoted to BVI, mainly in collaboration with the {\em Organizaci\'on Nacional de Ciegos de Espa\~{n}a} (ONCE), but also to all kind of publics at different educational levels, covering from primary to college, always using an inclusive methodology in any case. In all these activities it is common to resort to  complete oral descriptions of the presented material, in combination with adequate comparisons, highlighting the relative weight of the sense of sight in the acquisition of the explained information.

Another important resource used in all the activities of {\em Astroaccesible} is the tactile material, under the form  of both sheets in relief or 3D models, as those developed by the project {\em A Touch of the Universe} \cite{ortiz19}, representing some of the rocky planets and moons of the Solar System. 

In addition, 3D models of the night sky printed on the outer surface of an hemisphere, and representing stars in relief, with lines joining stars of the same constellations have been also widely used in our activities. These models help BVI to establish a spatial mental image of the distribution of these elements on the sky, but can also help people able to see the same figures projected onto a screen, to identify and to put them in the context of the whole sky. We have used in the last years 3D models representing the night sky visible from the Northern hemisphere during autumn and winter (i.e. from Orion to Gemini constellations), an epoch when public observations are not frequent due to the adverse weather conditions. For this reason, in this contribution, we present a new model representing the night sky during spring and summer, with most of the constellations observable during this period that can be used by both the collective of BVI and people without problems of vision during outdoors or indoors inclusive activities.  

\section{Description and goals of the 3D models \label{models}}

The 3D models described here consists of an hemisphere whose outer surface presents different elements in relief representing stars and deep sky objects  that can be observed in the night sky from the Northern hemisphere, at a latitude similar to that of southern Europe, during spring and summer.

The different elements in relief include points representing stars. The size of these points is proportional to the brightness of the corresponding star. Stars belonging to a same constellation are joined by
solid lines. The represented constellations include Virgo, Scorpius, Sagitarius, Lyra, Cygnus, Andromeda, Perseus, Ursa Major, and Ursa Minor.
These are also joined between them with dashed lines that serve as a guide to travel throughout the model using the sense of touch. In addition, a coarse band goes through the surface representing the Milky Way, as observed during the same epoch. Finally, there are some other coarse extended circles representing deep sky objects, such as the Virgo cluster, the Ring Nebula, or the Andromeda galaxy. This portion of the night sky is easily
observable from places without an excess of light  pollution during spring and summer, when most of public observations made with an outreach purpose are carried out. Therefore, this material
can supply a valuable additional source of information for people of the BVI collective and, at same time, an additional  source of information to identify the same objects to those able to see them directly in the sky.   

\begin{figure}[!h]
\center
\includegraphics[width=0.4\textwidth]{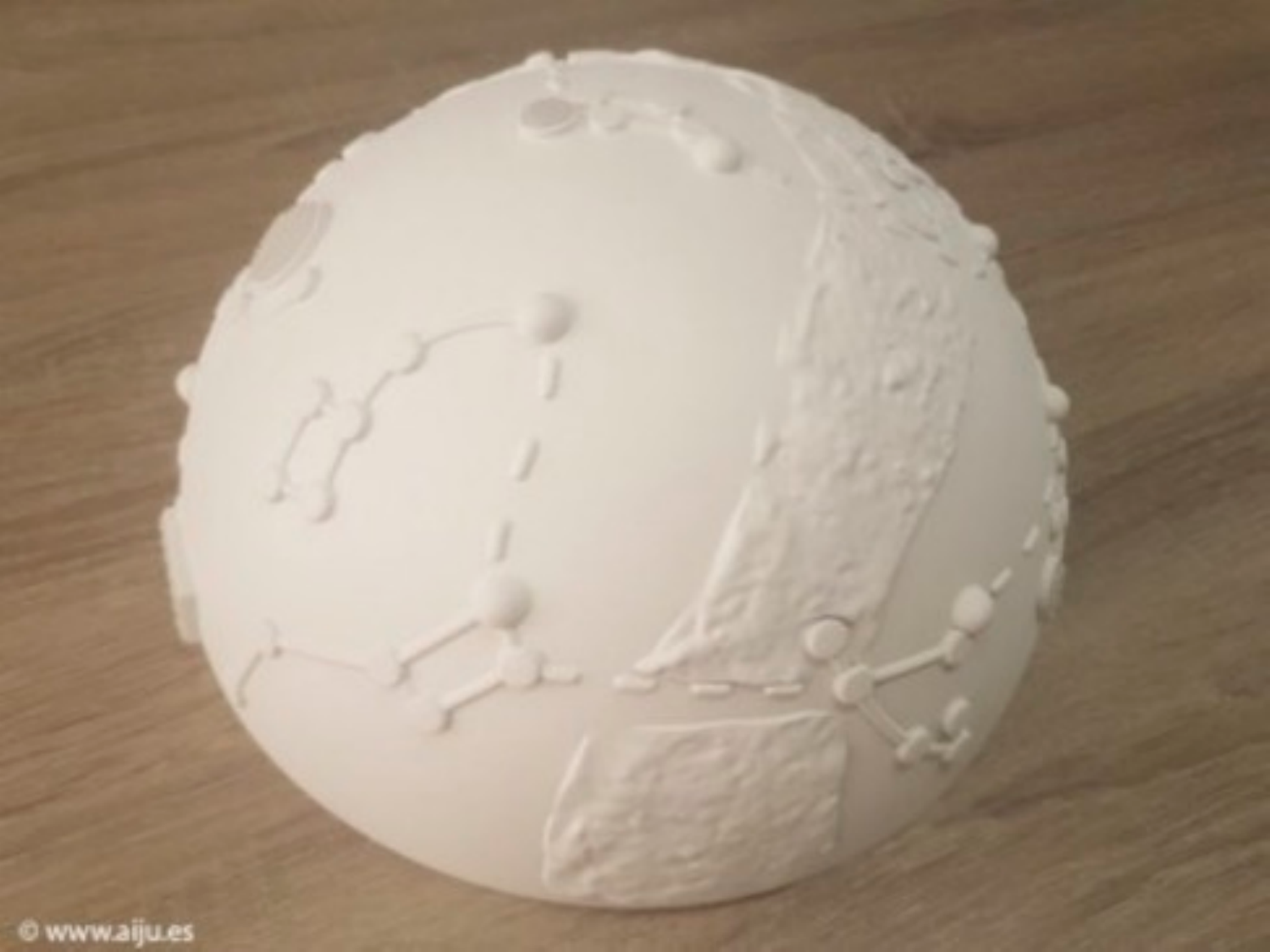} ~
\includegraphics[width=0.4\textwidth]{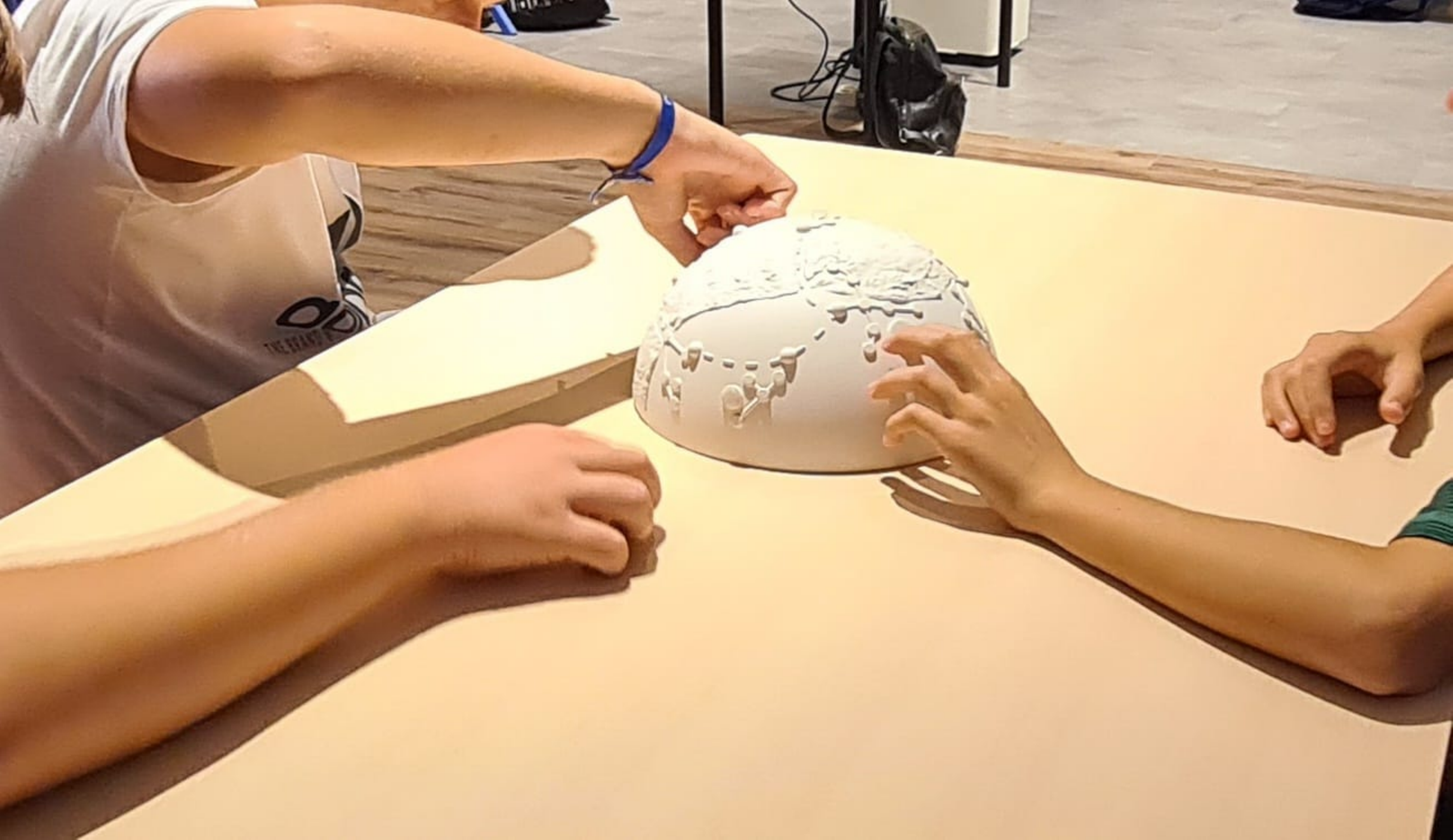} ~
\caption{\label{fig1} {\em Left:} Image of the printed 3D model representing the summer and spring northern night sky. {\em Right:} Image of one of  the activities carried out within the project {\em Astroaccesible} using the set of printed models. This activity was done in the {\em Museo Casa de la Ciencia} of CSIC in Sevilla (Spain) in October 2022 as part of the Space Week organized by this museum and was addresed to students of secondary level without any problem of vision.}
\end{figure}

A prototype using different elements in relief was designed, manually elaborated and
finally digitalized. 
From this, we printed several models that could be later used in face-to-face activities thanks to the financial support from the Sociedad Espa\~{n}ola de Astronom\'\i a (SEA) through its annual program of grants to outreach projects. An image of one of the printed resulting 3D models can be
seen in left panel of Figure \ref{fig1}. Additionally, the file necessary to create more copies of the same model, along with instructions describing the represented elements, both in English and Spanish, can be found \href{http://astroaccesible.iaa.es/content/descarga-el-cielo-de-verano-para-imprimir-en-3d}{in this link}.
The optimal diameter to print the models is 20 cm.

\section{Additional sonic resources \label{sounds}}

Sounds constitute an additional resource that can largely contribute to complement the multisensorial aspect of the
activities carried out by {\em Astroaccesible}, adding an extra dimension to all the treated material. The amount of available adaptations from astronomical data to sounds through their multiple techniques
(e.g. audification, sonification, or musification) helps to incorporate them to different adapted activities, as those related with the use of the models, deepen on their inclusive aspect.

For instance, among the different available resources, the {\em Cosmonic} project \cite{cosmonic}, whose different products can be consulted in its webpage \href{http://rgb.iaa.es/es/cosmonic/}{http://rgb.iaa.es/es/cosmonic/}, provides different animations with sounds, that allow the BVI to access to the data and simultaneously help people able to see the graphical content to properly interpret them.

An additional resource developed in the framework of {\em Astroaccesible} is the project {\em El Universo en palabras} ({\em The Universe in words}) \cite{euep}, a series of videos publicly available through a web channel in \href{https://www.youtube.com/playlist?list=PLDOpkwOM33-YAGnWAKSV9ZHiljg56o8mK}{YouTube} with images of different astronomical objects with an audio description in Spanish of both the context and the content of the image. Again, this resource can be useful for people of the BVI collective to overcome the barrier of the lack of information, supplying at same time a valuable background to everyone who want to better understand the described objects. 

Since many of the objects represented in the 3D model of the night summer sky can be also represented with either sonifications or audio descriptions, the simultaneous use of sounds in the activities focused on models ensures a much more efficiency in the teaching process when this material is used.

Moreover, the situation of pandemic suffered during most of 2020 and 2021, when most of in-person activities were cancelled and, therefore, the activities using tactile activities were not possible, the inclusive workshops participated by {\em Astroaccesible} were totally virtual and based exclusively on the use of sonic resources. This other modality of interaction, however, has also the advantage that it can be used for a more extended public involving participants from different geographical areas. Our experience from this period shows that a combination of both types of activities (i.e. in-person with models and virtual using only sounds) can be the best solution for the post-pandemic times.

\begin{figure}[!h]
\center
\includegraphics[width=0.5\textwidth]{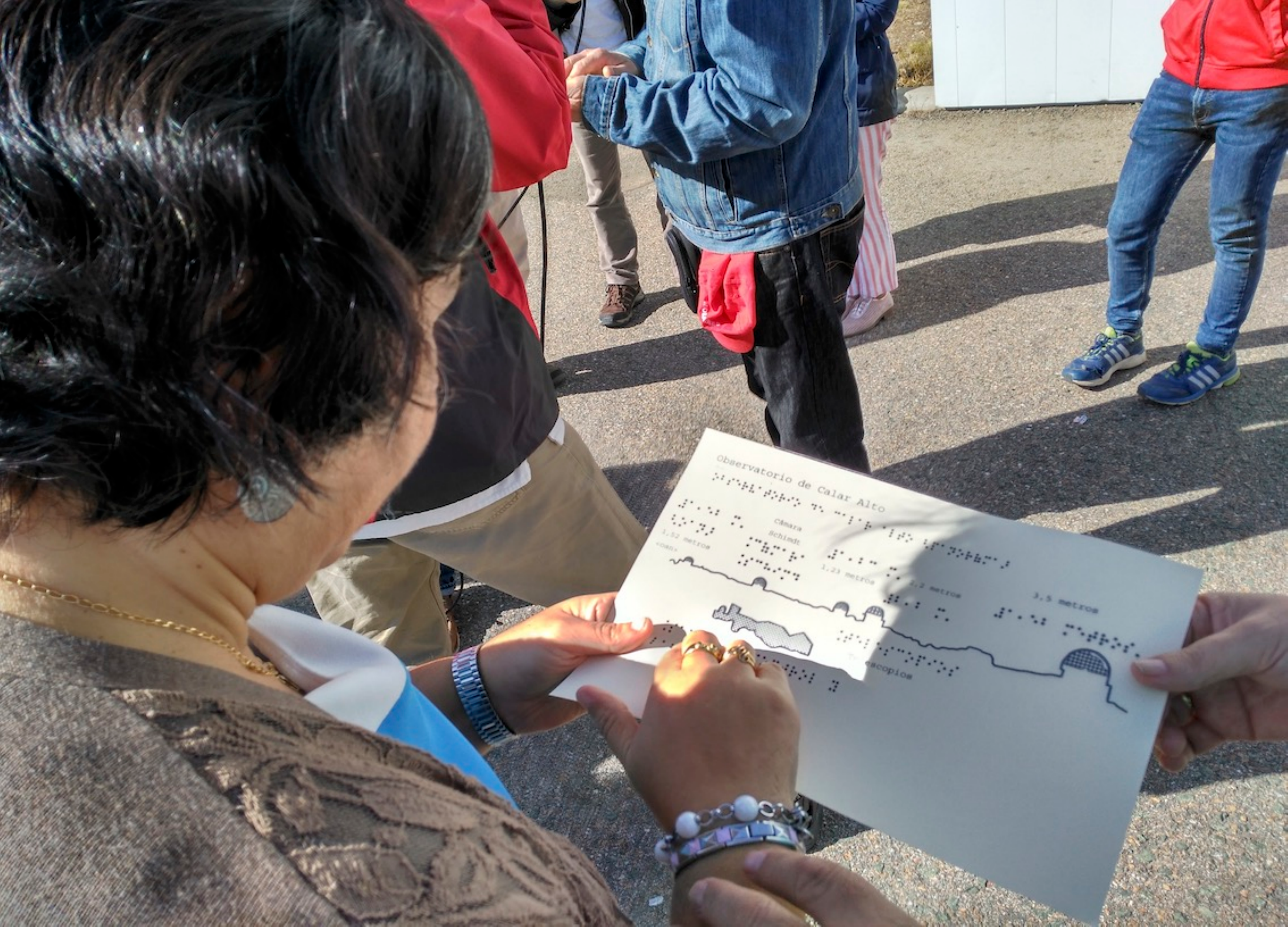} ~
\includegraphics[width=0.3\textwidth]{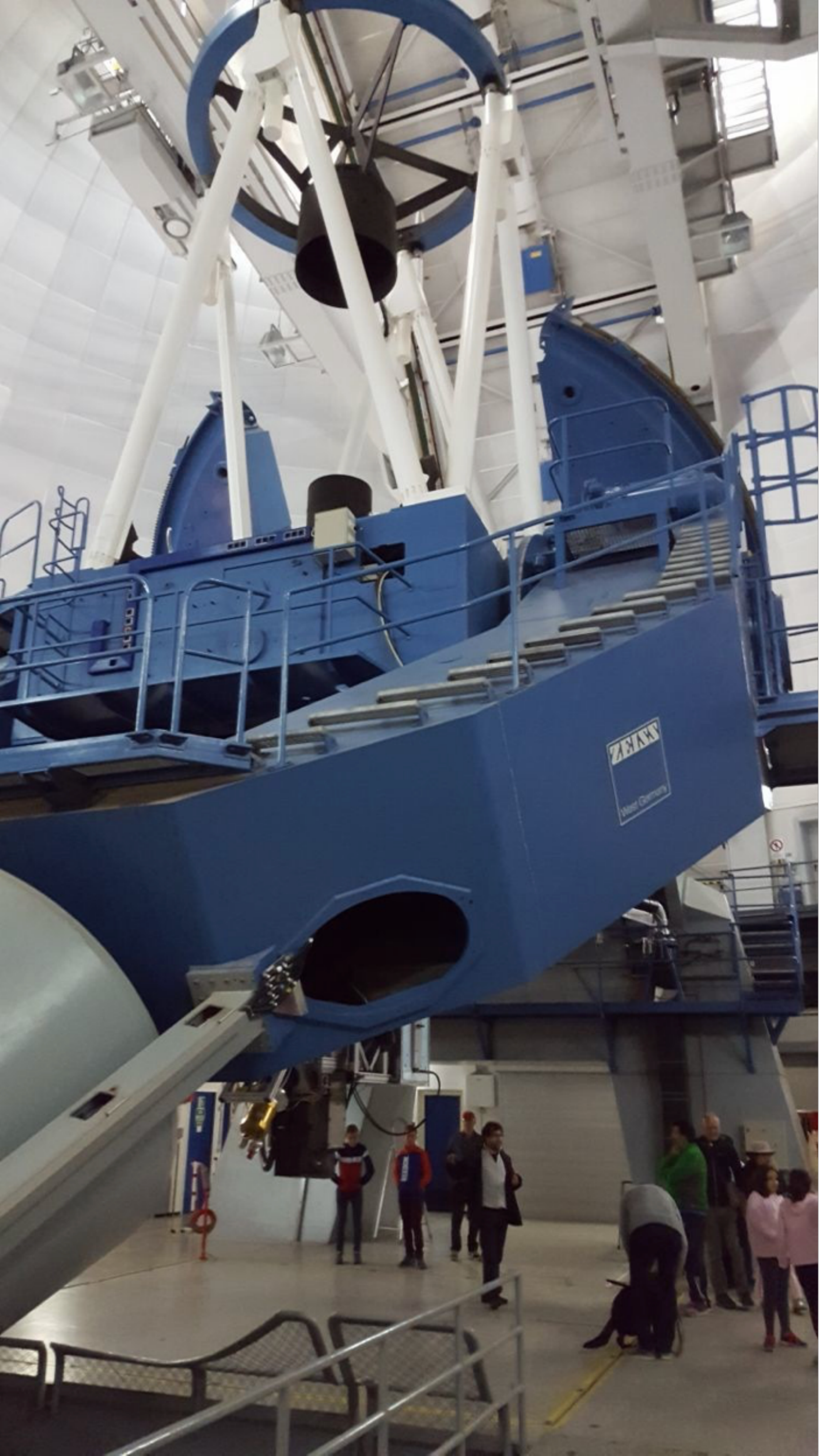} ~
\caption{\label{fig2} Images of one of the inclusive visits to the Calar alto Observatory {\em Left:} A person of the BVI collective exploring a sheet showing in relief some of the domes of the Observatory. {\em Right:} Picture of one of the groups under the dome of the telescope of 3.5 m.}
\end{figure}

 \section{Use of the 3D models in outreach activities}

Although the 3D models representing the summer night sky were designed and printed during 2019, due to the global situation of pandemic, they could not
be used in public outreach activities until the end of 2021. Fortunately, given the growing number of astronomical resources based on sounds, as explained in the previous section, allowed us to continue our inclusive activities addressed to an impaired and non-impaired public, during the period when in-person activities and the use of tactile material was restricted.
As an example, during April 2021 {\em Astroaccesible} took part in the organization of the {\em Month of Astronomy}, including several virtual activities addressed to members of ONCE in Spain. This program included virtual conferences using the sound as a resource to understand several astronomical objects, the public launch of the project {\em El Universo en palabras}, or the organization of a round table with several professional astronomers to describe without the use of images, different fields or problems of the astronomy. The conference using sounds was repeated for several educational institutions and scientific associations encouraging people to get interest about astronomy using sensorial channels complementing the sight.  

Once some in-person activities could be organized and the new 3D models could be used again, they were employed in scholar activities at different educational levels, such as the event {\em Noche Europea de los Investigadores}, or the {\em Aula Cient\'\i fica} in the University of Granada, with a high level of acceptance among the students and the teachers, even if these collectives did not include any participant with a visual disability. An example of these activities can be seen in right panel of Figure \ref{fig1} taken during the {\em Week of Space} organized by the {\em Casa Museo de la Ciencia} in Sevilla (Spain) for students of secondary level.

Among the other activities for which this set of models were used and exclusively addressed to people of the BVI collective, it is important to highlight the program of inclusive visits to the Observatory of Calar Alto (Almer\'\i a, Spain) financed by CSIC. We made three visits to this Observatory for BVI people in ONCE in the region of Andalusia, and another one for a intellectual disability association. The combined used of the models, sounds, images and inclusive explanations, along with the presence under the dome of several of the telescopes of the observatory was a complete experience that helped to fight against the idea that astronomy can only be learnt by people without any sight problem and reinforces the argument that multisensorial resources can be always applied to better transmit any scientific concept to any public. Two images of one of these inclusive visits are shown in Figure \ref{fig2}, one showing one person of the BVI collective exploring one sheet in relief showing the different domes present in the Observatory, and another one taken in the moment of the visit to the dome of the telescope of 3.5 m, the largest of the Observatory.

Therefore, the production of new tactile and sonic resources, not just in the outreach ambit, but also at a higher educational and research level can largely benefit the teaching process and the transmission of ideas and knowledge of all type of public and researchers.

%
%
\small  
%
\section*{Acknowledgments}   
%
We acknowledge the {\em Sociedad Espa\~{n}ola de Astronom\'\i a} (SEA) for its financial support to print most of the 3D models described in this proceeding, without its use the inclusive activities carried out by {\em Astroaccesible} would not be possible. We also thank to {\em Consejo Superior de Investigaciones Cient\'\i ficas} (CSIC) for its help to finance the inclusive visits to the Observatory of Calar Alto through its program {\em Cuenta la Ciencia}, during 2020.
EPM also thanks his guide dog Rocko, without whose daily help his both research and outreach activities would be much more difficult. 
%

%
\end{document}